\documentstyle[preprint,aps]{revtex}
\begin{document}

\draft

\title{Zero--Temperature Quantum Phase Transition of a Two--Dimensional
       Ising Spin--Glass}

\author{H. Rieger}
\address{Institut f\"ur Theoretische Physik, Universit\"at zu K\"oln,
50937 K\"oln, Germany}
\author{A. P. Young}
\address{Department of Physics, University of California, Santa Cruz, CA 95064}

\date{\today}

\maketitle

\begin{abstract}
We study
the quantum transition at $T=0$ in the spin-$\frac12$
Ising spin--glass in a transverse field in two dimensions.
The world line path integral representation of this model corresponds
to an effective classical system in (2+1)
dimensions, which we study by Monte Carlo simulations.
Values of the critical exponents are estimated by a
finite-size scaling analysis. We find that the dynamical exponent, $z$,
and the correlation length exponent, $\nu$, are given by $z = 1.5 \pm
0.05$ and $\nu = 1.0 \pm 0.1$.
Both the linear and non-linear susceptibility are found to diverge at
the critical point.
\end{abstract}


Much attention has been given to the {\em finite temperature} transition in
spin glass systems, see e.g.~\cite{review}, and reasonable agreement
between theory and experiment has been obtained.
This transition is
driven by {\em thermal} fluctuations controlled by the temperature. However, 
one can also control the strength of {\em quantum} fluctuations by
altering parameters in the system.
Turning up the quantum fluctuations will
decrease the transition temperature $T_c$, eventually
forcing it to zero. Critical fluctuations near the transition are
classical as long as $T_c > 0$, because they occur at a frequency
$\omega$ satisfying $\hbar \omega \ll k_B T$~\cite{slowingdown}.
Consequently, the universality class is that of
the classical problem except if
one tunes through the transition at $T = 0$.
This quantum universality class 
has not been much studied for the spin glass problem,
though other quantum phase transitions, such as the metal--insulator~\cite{mit}
and bose--glass~\cite{bg} transitions,
have attracted a lot of attention. Most theoretical work on the quantum
spin glass~\cite{mft1,mft2} has been confined to the
infinite range model, which is expected to describe the transition in
a short range system of sufficiently high space dimension.

Recently, however, the quantum spin glass transition was studied
experimentally~\cite{exp}
in an Ising system with dipolar couplings in which $T_c$ was driven to
zero by applying an effective transverse field. Interestingly, the
non-linear susceptibility, $\chi_{nl}$, which diverges at the
finite-$T$ classical transition~\cite{review}, was found not to diverge,
or at least to diverge much less strongly than in the classical case.
Furthermore, the phase transition in a
quantum Ising spin system in (1+1) dimensions has recently been studied in
detail~\cite{DF}, see also~\cite{ising}.
It is found that both the linear and non-linear
susceptibility diverge not only at the critical point but also in part
of the disordered phase. Although this model does not have frustration,
and therefore might miss some of the spin glass physics, it is 
interesting to investigate whether similar behavior also occurs
higher dimensions. It is therefore an appropriate time to study
the quantum Ising spin glass and
here we report on results of Monte Carlo
simulations on a short range model in (2+1) dimensions.
Similar calculations and analysis
have also been performed in (3+1) dimensions~\cite{bhahus}.

The model system studied in this paper, which  is
appropriate for the experimental system, LiHo$_x$Y$_{1-x}$F$_4$~\cite{exp}, is
the Ising spin glass in a transverse field with Hamiltonian
\begin{equation}
H=-\sum_{\langle ij\rangle}J_{ij}\sigma_i^z\sigma_j^z-
\Gamma\sum_i\sigma_i^x\;,\label{eq1}
\end{equation}
where the $\sigma_i$ are Pauli spin matrices, $\Gamma$ is the strength
of the transverse field and the
nearest neighbor interactions, $J_{ij}$,
are independent random variables with
a Gaussian distribution of mean zero and standard deviation unity.

If $\Gamma=0$ the
Hamiltonian in (\ref{eq1}) is the 
classical two dimensional Ising spin--glass. The ground state is doubly
degenerate (the two states being related by global spin--flip symmetry) so,
at $T=0$,
the Edwards--Anderson order parameter \cite{review}
$q_{EA}
=[\langle\sigma_i^z\rangle^2]_{\rm av}$ is unity.
We denote a statistical 
mechanics average by angular brackets,
$\langle\cdots\rangle$, and an average over the quenched disorder by
square brackets, $[\cdots]_{\rm av}$.
Switching on the transverse field mixes the eigenstates of $\sigma^z$
and thus diminishes the EA--order parameter, causing it to vanish
at some finite value, $\Gamma_c$. This is the transition that we study
here.
Details of the calculations will be given elsewhere \cite{big}.

It is well known \cite{suzuki} that the ground state energy of the
$d$--dimensional quantum mechanical model (\ref{eq1}) is equal to the
free energy of a $(d+1)$--dimensional classical model, where the extra
dimension corresponds to imaginary time, i.e.
\begin{equation}
-{ E( T = 0) \over L^d} = \lim_{T\to 0} {T \over L^d} {\rm Tr\;} e^{-\beta H}
=
{1 \over \Delta\tau}
{1 \over L_\tau L^d} {\rm Tr\;}  e^{-{\cal S}}
\end{equation}
where the
imaginary time direction has been divided into $L_\tau$ time slices of
width $\Delta\tau$ ($\Delta\tau L_\tau = \beta$), and
the effective classical action, $\cal S$, is given by
\begin{equation}
{\cal S} =
-\sum_{\tau}\sum_{\langle ij\rangle}
K_{ij} S_{i}(\tau) S_{j}(\tau)
-\sum_{\tau}\sum_i K S_{i}(\tau) S_{i}(\tau+1)\;,
\label{eq2}
\end{equation}
where the $S_i(\tau)=\pm1$ are classical Ising spins, the indices $i$ and $j$
run over the sites of the original $d$--dimensional lattice
and $\tau = 1,2,\ldots,L_\tau$ denotes a time slice. In Eq. (\ref{eq2}), 
$K_{ij} = \Delta\tau J_{ij} $ and $\exp(-2 K) = \tanh(\Delta\tau \Gamma)$.
Note that we have the {\em same} random interactions in each time slice.
We should take the limit
$\Delta\tau \to 0$, which implies
$K_{ij} \to 0$ and $K \to \infty$. This extremely
anisotropic limit is inconvenient for calculations but
universal properties are
expected to be independent of $\Delta\tau$ so we take
$\Delta\tau = 1$ and set the standard deviation of
the $K_{ij}$ to equal $K$. Thus $K$, which physically
sets the relative strength of the
transverse field and exchange terms in (\ref{eq1}), is like an inverse
``temperature'' for the effective classical model in (\ref{eq2}).

We study the model (\ref{eq2}) in $d=2$ dimensions by Monte--Carlo 
simulations on a simple cubic lattice of size 
$L\times L\times L_\tau$ using periodic boundary conditions.
Since various quantities
of interest show a very strong dependence on the disorder realization we
have to average over a large number of samples --- we took 2560 samples for
each temperature and size. The largest systems were 
$20\times20\times50$, where we used up to $10^5$ Monte Carlo sweeps
for equilibration plus $10^5$ sweeps for measurements, which were
performed every
20 sweeps. Equilibration was checked with standard methods 
\cite{bhayou}. The simulations were performed on a 
large transputer array (GCel1024 from Parsytec).

Because the system in 
(\ref{eq2}) is very anisotropic, it is expected to
have two different diverging scales: one is the correlation length
in the space direction, $\xi\sim \delta^{-\nu}$, where $\delta=K_c/K - 1$
is the distance from
the critical point $K_c$, and the other is the correlation time,
$\xi_\tau$, in the
(imaginary) time direction, where
$\xi_\tau\sim\xi^{z}$ with $z$ the dynamical
exponent.
According 
to a finite size scaling hypothesis extended to anisotropic systems 
\cite{binder}, various thermodynamic quantities close to the critical point 
depend on {\it two} independent scaling variables, which we can take to
be $\delta L^{1/\nu}$ and the aspect ratio $L_\tau/L^z$. The scaling
analysis is straightforward only if
it depends on a single parameter, so it is necessary to fix the aspect ratio.
Since $z$ is unknown,
one has to scan several
different sample shapes to see which choice for $z$ scales best, and
we follow an efficient method of doing this suggested by Huse~\cite{huse}.

As in standard spin--glass theory \cite{review}, we define the overlap 
between the configurations of two replicas, 1 and 2, with the same
disorder as
\begin{equation}
Q= {1 \over L^d L_\tau} \sum_{i,\tau}S_i^{(1)}(\tau) S_i^{(2)}(\tau) \quad ,
\label{overlap}
\end{equation}
and for each disorder realization we calculate the dimensionless combination
of moments
\begin{equation}
g=0.5 \Bigl[3-\langle Q^4\rangle/\langle Q^2\rangle^2 \Bigr] \quad .
\label{g:def}
\end{equation}
The disorder averaged quantity, $g_{\rm av}=[g]_{\rm av}$ \cite{remark}, 
obeys the finite size scaling form
\begin{equation}
g_{\rm av}(K,L,L_\tau)={\tilde g}_{\rm av}(\ \delta L^{1/\nu},\ L_\tau/L^z\ )\;,
\label{eq3}
\end{equation}
and has the property~\cite{bhayou}
that it vanishes in the disordered phase for $L\to \infty$,
and tends to a finite value in the ordered phase. Consequently, 
$\tilde{g}(x,y)$ vanishes at fixed $x$ both for $y\rightarrow0$
(where
the system is a classical two--dimensional spin glass at finite
``temperature'', which is disordered)
as well as for $y\rightarrow\infty$
(where the system is effectively a long
one--dimensional chain along the $\tau$ direction, which is also
disordered).
Hence, $\tilde{g}(x,y)$ must have a maximum at some value of $y$ for
fixed $x$. The value of
this maximum decreases with increasing $L$ in the disordered phase $K <
K_c$ (where $\delta = (K_c / K - 1) > 0$) and increases with increasing $L$
in the ordered phase. We use this criterion to estimate the critical
coupling which we find is given by
$K_c^{-1}=3.275 \pm 0.025$. The data are shown in Fig.\ 1.
Furthermore, at the critical point, the values of $L$ and $L_\tau$ for
which $\tilde{g}$ is a maximum are related by $L_\tau \sim L^z$.
By this method we determine the dynamical
exponent and get $z=1.50\pm0.05$. The finite--size scaling hypothesis
(\ref{eq3}) can be checked a posteriori by a scaling plot for $g_{\rm av}$
at $K_c$ as shown in Fig.\ 2.

Systems with fixed aspect ratio, $L_\tau / L^z$,
can be used them to determine critical exponents
via the usual one--parameter finite--size scaling. First of all,
from Eq.~(\ref{eq3}) the
derivative of $\tilde{g}$ with respect to $K$ at $K_c$ gives $\nu$ and
we find $\nu = 1.0 \pm 0.1$, see Fig. 3.
The rigorous inequality $\nu \ge 2 / d $~\cite{chayes} is therefore
satisfied, perhaps as an equality.

There are various susceptibilities that one can define for this problem,
with different numbers of integrations over imaginary time.
For example, the second moment of Q,
$ \chi_Q = L^d L_\tau [ \langle Q^2 \rangle ]_{\rm av} ,$
has a single integral over $\tau$. Defining the exponent $\gamma_Q$ by
$\chi_Q \sim \delta^{-\gamma_Q}$, then, at the critical point, the size
dependence is given by $\chi_Q \sim L^{2-\eta}$ where $\gamma_Q = (2 -
\eta) \nu$.
On the other hand, the equal time spin glass
correlation function, $C_0 = \sum_i [ \langle  S_{i_0}(\tau_0) S_i(\tau_0)
\rangle^2 ]_{\rm av} $,
has no $\tau$ sum and so varies as
$L^{2 - \eta - z}$~\cite{remark4}.
Consider next the overlap
\begin{equation}
q^{ab}= {1 \over L^d L_\tau^2} \sum_{i,\tau_1,\tau_2}S_i^{(a)}(\tau_1)
S_i^{(b)}(\tau_2) \quad .
\label{qdef}
\end{equation}
which involves a double sum over $\tau$. The corresponding susceptibility,
$\chi_q = L^d L_\tau^2\ [ \langle\ (q^{12})^2 \ \rangle ]_{\rm av} ,$
involves {\em two} time integrals so it should vary as $L^{2 - \eta + z}$ at
criticality~\cite{remark4}. The experimentally measured
non-linear susceptibility is the fourth derivative of the free energy
with respect to a field coupling to $S^z$, and so 
is related to the fourth order
cumulant of the total magnetization by standard linear response theory,
$\chi_{nl} = \left[ \langle M^4 \rangle - 3 \langle M^2 \rangle^2
\right]_{\rm av} / (L^d L_\tau) , $
where $M = \sum_{i,\tau} S_i(\tau)$.
Since the disorder average gives zero unless each spin
occurs an even number of times, $\chi_{nl}$ can be expressed (neglecting a
local piece which diverges less strongly) as
\begin{equation}
\chi_{nl} = L^d L_\tau^3 \left[ \langle (q^{12})^2 \rangle -
\frac{1}{4} \langle (q^{11} - q^{22})^2 \rangle \right]_{\rm av}
\label{chinl}
\end{equation}
which has {\em three} sums over $\tau$ and so should diverge at criticality
like $L^{2 - \eta + 2z}$~\cite{remark4}.
Fig. 4 shows data and fits
for $C_0, \chi_q, \chi_Q$ and $\chi_{nl}$ at criticality. All the data
are consistent with the exponent values, $\eta \simeq 0.5, z \simeq
1.5$. In particular, 
$\chi_{nl} \sim L^{4.7}$ at criticality, or equivalently
$\chi_{nl} \sim L_\tau^{3.1}$ using $z = 1.5$. Since $L_\tau \propto \beta$,
$\chi_{nl}$ varies as $T^{-3.1}$ for $T \to 0$ at the critical
transverse field $\Gamma_c$,
which is quite a strong divergence. Note that, by contrast, the equal
time correlation function
does not diverge (or only does so marginally). This
is because spatial correlations fall off quite rapidly
at criticality, like $r^{-2}$, as we have verified directly.

According to scaling theory~\cite{bg}, the (unsquared) on-site correlation
function at the critical point, $C(\tau) = [\langle S_i(0) S_i(\tau)
\rangle ]_{\rm av}$ varies as $\tau^{-(d+z - 2 + \eta)/(2z)}$, or
$\tau^{-2/3}$ using our values for the exponents. Integrating this over
$\tau$ to get the uniform susceptibility, $\chi_F$~\cite{remark3}, 
$\chi_F = \sum_\tau C(\tau)$, one finds a divergence of the form
$L_\tau^{1/3}$, or $\chi_F \sim
T^{-1/3}$ as $T\to 0$ at $\Gamma = \Gamma_c$.
Thus, in contrast
to the classical spin glass~\cite{review}, the
uniform susceptibility diverges at the quantum spin glass transition in
(2+1) dimensions. 

Similar calculations and analysis have been performed on a (3+1)
dimensional model~\cite{bhahus}, with results which are quite
similar to ours, though the numerical
values for exponents are somewhat different as expected.
The main qualitative difference is that the
uniform susceptibility does not diverge in (3+1) dimensions.
Both our work and the results in (3+1)
dimensions~\cite{bhahus} show a substantial divergence of $\chi_{nl}$,
which appears to be rather different from experiment~\cite{exp}. The
reason for this discrepancy is unclear at present.
For future work
it will be interesting to investigate whether the uniform and spin glass
susceptibilities diverge in part of the disordered phase, as
happens in $d =$ (1+1) because of Griffiths singularities arising from rare
regions which are more strongly coupled than the average~\cite {DF}.

We would like to thank R.\ Bhatt and D.\ Huse for many
extremely valuable suggestions and discussions. We also
thank N.~Read for a helpful conversation. HR acknowledges
the kind hospitality of the Physics Department of UCSC, where this
work was begun, as well as
the generous allocation of computing time on the transputer cluster
Parsytec--GCel1024 from the Center of Parallel Computing (ZPR)
in K\"oln and the
friendly user--support by M.\ Wottawa. APY is supported by
NSF DMR-91-11576, and HR's work was performed within the SFB 341
K\"oln--Aachen--J\"ulich.

\begin{figure}
\caption{The averaged cumulant $g_{\rm av}(K,L,L_\tau)$ for
three different coupling
constants ($K^{-1}$=3.20 left, $K^{-1}$=3.30 middle and
$K^{-1}$=3.40 right) and various systems sizes (L=4 ($\diamond$), L=6
($+$), L=8 ($\Box$), L=12 ($\times$) and L=16 ($\triangle$)) as a
function of $L_\tau$. The maximum increases with $L$
for $K^{-1}$=3.20, which implies $K_c^{-1}>3.20$,
and it decreases with increasing
$L$ for $K^{-1}=3.40$, so $K^{-1}_c<3.40$.
We have also data for $K^{-1}$=3.25,
from which we conclude that $K^{-1}_c$ is between 3.25 and 3.30.
The errorbars are smaller than the symbols.}
\end{figure}

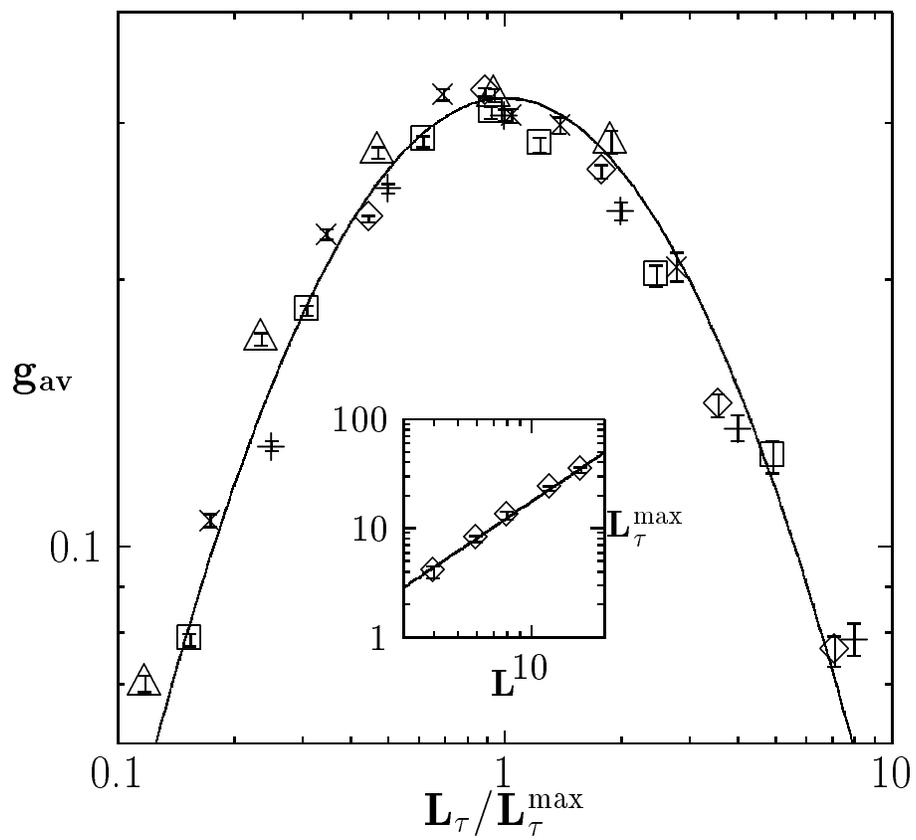
\begin{figure}
\caption{A scaling plot of $g_{\rm av}(K,L,L_\tau)$
at $K^{-1}=3.30 \simeq K_c^{-1}$ as a function of the
scaled system size in the (imaginary) time direction
$L_\tau/L_\tau^{\rm max}$.
For each lattice size, $L_\tau^{\rm max}$ is chosen so that all the data 
collapses on to a single curve.
The sizes are
(L=4 ($\diamond$), L=6($+$), L=8 ($\Box$), L=12 ($\times$) and
L=16 ($\triangle$)).
The inset shows the dependence of $L_\tau^{\rm max}$ as a function
of $L$. From Eq.~(\protect\ref{eq3}) the slope is equal to the
dynamical exponent $z$ and a fit gives $z= 1.50\pm0.05$.}
\end{figure}

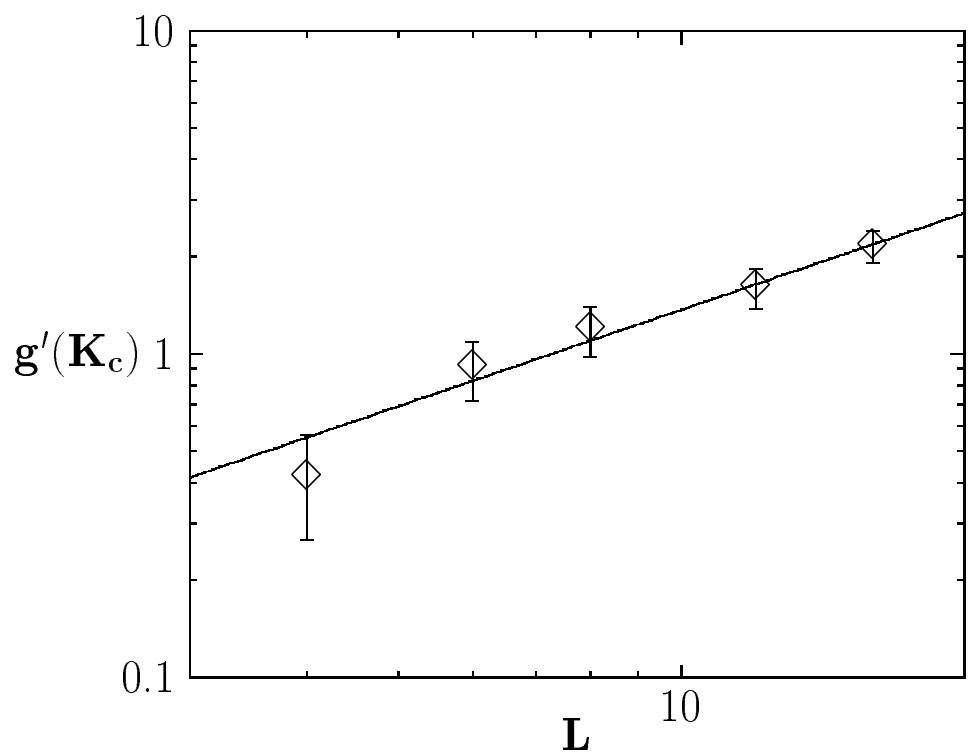
\begin{figure}
\caption{
The derivative of $g_{av}$ with respect to $K^{-1}$
at $K^{-1} = 3.30 \simeq K^{-1}_c$, for systems
of size $4\times4\times4$, $6\times6\times8$, $8\times8\times14$,
$12\times12\times24$ and $16\times16\times34$, which have a roughly
constant aspect ratio, $L_\tau / L^z$, since $z \simeq 1.5$.
A least squares fit of the data to a straight line
yields a slope of $1/\nu=1.0\pm0.1$.}
\end{figure}

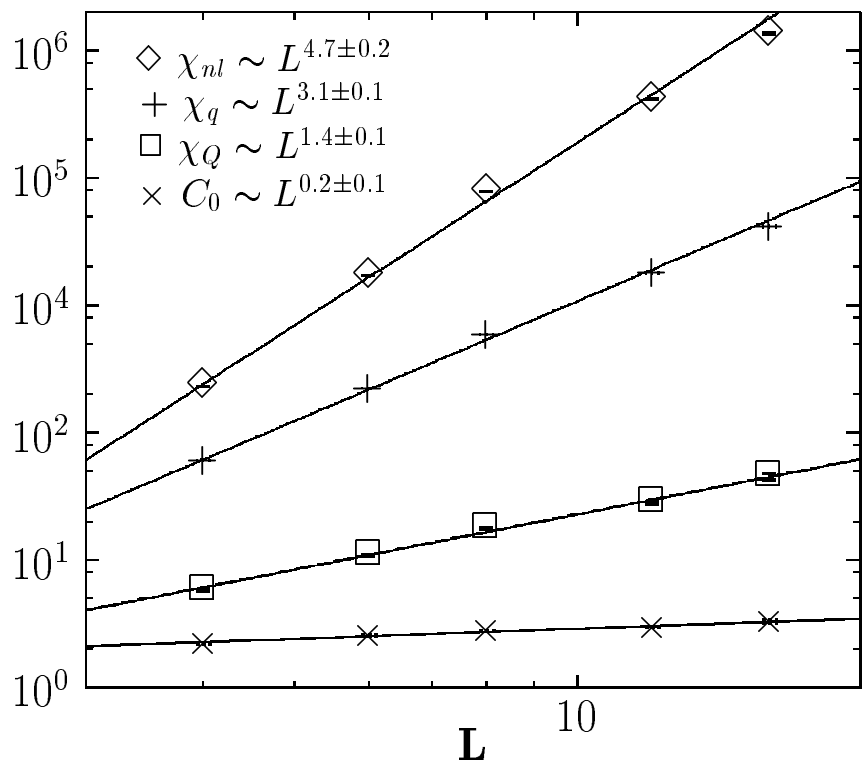
\begin{figure}
\caption{
The equal time correlation function, $C_0$, and the susceptibilites
$\chi_Q$, $\chi_q$ and $\chi_{nl}$ as a function of $L$ close to the critical
point, $K^{-1} = 3.30 \simeq K_c^{-1}$,
on a double logarithmic plot. The slopes are expected
to be $2 - \eta - z$, $2 -\eta$, $2 -\eta + z$, and $2 - \eta + 2z$,
respectively. A least squares fit gives the values $0.2 \pm 0.1$,
$1.4\pm 0.1$, $3.1 \pm 0.1$ and $4.7 \pm 0.2$, which are consistent with
the exponents, $\eta \simeq 0.5, z \simeq 1.5$.  The system sizes are
are the same as in Fig.~3.}
\end{figure}


\begin{references}
\bibitem{slowingdown}
Because of critical slowing down, the characteristic frequency
tends to zero at the critical point.
\bibitem{huse}
We would like to thank David Huse for very enlightening
discussions on this point.
\bibitem{remark}
We also tried averaging the numerator and denominator in Eq.~(\ref{g:def}) 
separately over the disorder, as was done in the classical spin 
glass~\cite{bhayou}. An analysis gives similar results~\cite{big}
but with worse
statistics because of strong sample--to--sample fluctuations.
\bibitem{remark4}
The divergence of the various susceptibilities can also
be obtained by noting that $Q$ (and $q$ defined in Eq.~(\protect\ref{qdef}))
is the order
parameter for this problem and so, according to finite-size scaling,
both $[ \langle Q^2 \rangle ]_{\rm av}$ and
$[ \langle q^2 \rangle ]_{\rm av}$ vary as $L^{-2 \beta / \nu}$ at
criticality. Note also that the equal time correlation function $C_0$
is of the form $L^d$ times the square of the order parameter.
In addition, $L_\tau \propto L^z$, and, according to
hyperscaling with $d$ replaced by $d +z$~\cite{bg}, one has
$2\beta / \nu = d + z - 2 +\eta$. These considerations then lead to the 
exponents summarized in the caption to Fig.~4.
\bibitem{remark3}
Contributions to $\chi_F$
of the type $[ \langle S_i(0) S_j(\tau)\rangle ]_{\rm av}$, 
where $i \ne j$, vanish for a symmetric bond distribution so the uniform
susceptibility is equal to the local (on--site) susceptibility.
\end{references}
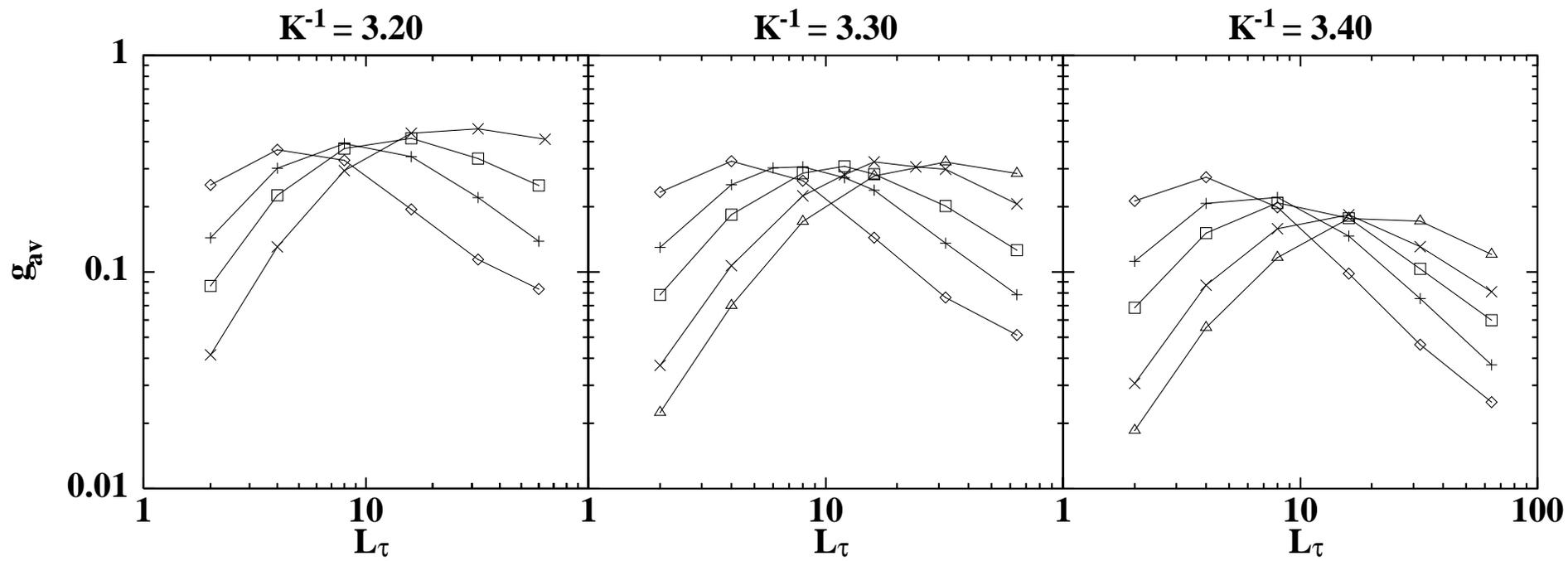
\end{document}